Corresponding author information

Corresponding Author:

Jonathan F. Donges, Potsdam Institute for Climate Impact Research, Telegrafenberg A31, 14473 Potsdam, Germany and Stockholm Resilience Centre, Stockholm University, Kräftriket 2B, 114 19 Stockholm, Sweden

Email: donges@pik-potsdam.de

# Title: The technosphere in Earth system analysis: a coevolutionary perspective

## Authors

Jonathan F. Donges[1,2]. Wolfgang Lucht[1,3]. Finn Müller-Hansen[1,4]. Will Steffen[2,5]

[1]Potsdam Institute for Climate Impact Research, Potsdam, Germany

[2]Stockholm Resilience Centre, Stockholm University, Stockholm, Sweden

[3]Department of Geography, Humboldt University, Berlin, Germany

[4]Department of Physics, Humboldt University, Berlin, Germany

[5]Fenner School of Environment and Society, The Australian National University, Canberra, Australia

## Abstract

Abstract

Earth system analysis is the study of the joint dynamics of biogeophysical, social and technological processes on our planet. To advance our understanding of possible future development pathways and identify management options for navigating to safe operating spaces while avoiding undesirable domains, computer models of the Earth system are developed and applied. These models hardly represent dynamical properties of technological processes despite their great planetary-scale influence on the biogeophysical components of the Earth system and the associated risks for human societies posed, e.g., by climatic change or novel entities. In this contribution, we reflect on the technosphere from the perspective of Earth system analysis with a threefold focus on agency, networks and complex coevolutionary dynamics. First, we argue that Haff's conception of the technosphere takes an extreme position in implying a strongly constrained human agency in the Earth system. Assuming that the technosphere develops according to dynamics largely independently of human intentions, Haff's perspective appears incompatible with a humanistic view that underlies the sustainability discourse at large and, more specifically, current frameworks such as UN sustainable development goals and the safe and just operating space for humanity. Second, as an alternative to Haff's static three-stratum picture, we propose complex adaptive networks as a concept for describing the interplay of social agents and technospheric entities and their emergent dynamics for Earth system analysis. Third, we argue that following a coevolutionary approach in conceptualising and modelling technospheric dynamics, also including the socio-cultural and biophysical spheres of the Earth system, could resolve the apparent conflict between the discourses on sustainability and the technosphere. Hence, this coevolutionary approach may point the way forward in modelling technological influences in the Earth system and may lead to a considerably deeper understanding of pathways to sustainable development in the future.

# Introduction

As a defining characteristic of the Anthropocene, human societies have created large-scale technological infrastructures such as world-spanning industrialized energy and food production and distribution systems for supporting historically unprecedented numbers of human beings embedded in increasingly complex socio-cultural structures while significantly intervening in the dynamics of the Earth system on a planetary scale. In this way, the worldwide evolving network of mutually interdependent technological and social macrostructures (examples for the latter include modern states, bureaucracies and social institutions in general), the *technosphere* in the sense of Haff (2014a), gives rise to key global environmental crises. These crises and their local and regional manifestations are reflected in the transgression of planetary boundaries such as those related to anthropogenic climate change, degenerative land-use change, accelerated biodiversity loss, perturbation of the global biogeochemical cycles of nitrogen and phosphorus, and the creation and release of *novel entities* such as nanoparticles and genetically engineered organisms (Steffen et al., 2015b).

In this contribution to the Anthropocene Review's Special Issue on the technosphere, we reflect on the implications and relevance of Haff's concept in the context of Earth system analysis. This field of research explores possible future development pathways compatible with the coevolutionary dynamics of the biogeophysical and socio-technological spheres and aims at identifying management options for navigating to sustainable safe operating spaces while avoiding undesirable Earth system states such as "catastrophe domains" (Schellnhuber, 1998; 1999). Our contribution intends to connect separate discourses about the technosphere on the one hand and Earth system analysis and sustainable development on the other hand by providing insights into current debates on how to include technological dynamics in Earth system models and exploring how the concept of the technosphere could be used to advance the understanding of these dynamics. We begin by discussing human agency in Haff's technosphere concept from the perspective of sustainability science. Then we briefly consider the relevant state-of-the-art of modelling technological dynamics in Earth system science and discuss issues of collective human agency in this context. Finally, we propose a complex systems approach for analytically dealing with the technosphere in the Earth system that is founded on (i) coevolutionary dynamics and emergence and (ii) adaptive Earth system networks.

# The technosphere and human agency

Agency is a key concept in the Anthropocene discourse. It arises as a crucial issue when considering an Earth system that is not only influenced by a socio-technological complex but also generates with increasing severity unintended consequences from the actions of that complex with repercussions for human societies. The notion of agency is traditionally debated in philosophy and sociology, but has received much attention as well in psychology and neuroscience in the last decades. Put simply, in these fields agency is the human experience of being the subject or owner of one's actions. This sense of agency is refined in the philosophy of action, where the term usually refers to the capability of an agent to perform deliberate and intentional action as opposed to forced, determined or random behaviour (Schlosser, 2015; Moya, 1990). The sociological concept of agency is often used as an antonym to social structure (Elder-Vass, 2010; Ritzer, 2010). On the one hand, structure determines the individual's actions and behaviour. On the other hand, structure emerges from the actions of individuals, forming a coevolutionary loop (Snijders et al., 2010). The concept of agency of the individual emphasizes some degree of potential primacy of the individual over structure. Thus agency can be understood as one part of a dialectic understanding of the social.

Haff's concept of the technosphere shifts the focus from social relations to relations between humans and technology, a theme that is explored from other perspectives in the field of science and technology studies (e.g., Latour, 2007). Haff raises important questions regarding human agency and the controllability of large-scale technologies as well as the role of technology in the interrelation between human societies and other parts of the Earth system. Haff attempts to take a physicist's outside point of view on the technosphere as a "geological phenomenon", postulating that the technosphere follows some "physical law" or "quasi-autonomous dynamics" such as the principle of maximum entropy production (Haff, 2014a). From this perspective, human agency and purpose may have been the originators of technological systems, but are no longer their controlling factor. Haff thus presents an account of recent human development as only a part of the systemic dynamics of the technosphere, thereby challenging the intuition that political decisions and societal change are solely the result of human volitions.

Haff notes that human actions are strongly constrained by technological possibilities and dependencies. Technologies and institutions increase societies' robustness to external and internal disturbances but also constitute path-dependencies and lock-ins that make large-scale changes difficult. The energy system is a good example of such a lock-in: Investments in fossil fuel technology can be considered as costs that owners of such investments wish to recover (and large parts of society wish to make use of). A radical shift in energy production towards renewable energies would make these prior investments worthless. Thus, a rapid transition to renewable energies is proving to be difficult.

Haff takes this argument to the extreme: Motivated by an apparent separation of scales between the level of the individual and the large-scale technological complexes, as suggested by Haff's rules of inaccessibility and impotence (Haff, 2014a), humans as individuals do not in his view exert direct influence on the dynamics of the technosphere and hence its repercussions (Haff, 2014b). Similarly, other authors argue that social metabolism can be described as a thermodynamic machine with intrinsic momentum originating in the flows of energy and material required to construct, maintain and transform large-scale infrastructures (Garrett, 2014; 2015). Haff puts this extreme position only partly into perspective, by focussing on leadership and control. Even if humans might have agency on an individual level, he argues that they do not have it on the aggregate level. Instead, the argument in Haff's papers suggests that the technosphere has non-human agency which is in line with the discourse on the possibility of the emergence of general artificial intelligence and its consequences (Bostrom, 2014). The technosphere is presented as an emergent super-organism with its own teleology, desires and needs (Haff, 2014a), rather than serving human needs and normative goals.

Let us follow, for a moment, the assertion that the technosphere follows its own independent dynamics. This would imply that there is little room for political or ethical choice on a planetary level, e.g. for an intentional shift of technology towards sustainable production. Without the ability to influence technological development at a large scale, efforts to establish and implement normative goals such as UN sustainable development goals (Griggs et al., 2013) and frameworks such as sustainability paradigms (Schellnhuber, 1998; 1999) and the safe and just operating space for humanity (Rockström et al., 2009; Raworth, 2012) would be futile. Taking this into consideration, Haff's concept of the technosphere is incompatible with a sustainable development discourse founded on humanitarian principles.

However, we currently see at best mixed evidence for technology following its own dynamics in a manner that is totally independent of human intentions. Certainly, deviations from the business-as-usual path require more effort to succeed because well-established vested interests and technological inertia have to be overcome. But there is, in our view, no *a priori* reason why normative goals are unachievable; there are many examples where policies can opt out of large-scale technologies (e.g. the global banning of CFCs and the nuclear fade out in Germany). Instead of taking the development of technology as given, we suggest to do the opposite: frame it as a political question, i.e. regarding collective social action.

This is perhaps the most fundamental shortcoming of Haff's technosphere concept as it stands: the fact that humans reflect on their relationship to the world and adapt their actions accordingly does not seem to have any consequence for the emergent phenomenon of the technosphere. But history, for example of economic institutions, shows that theories about human societies and their environments can influence their behaviour, sometimes even leading to situations of self-fulfilling or self-defeating prophecy (e.g. Ferraro et al. 2005). Therefore, we think it is essential to consider human reflexivity as an integral part in the coevolution between technology and human societies.

In the following, we aim for a more differentiated understanding on the technosphere concept building on Haff's notion that "The technosphere includes the world's large-scale energy and resource extraction systems, power generation and transmission systems, communication, transportation, financial and other networks, governments and bureaucracies, cities, factories, farms and myriad other 'built' systems, as well as all the parts of these systems, including computers, windows, tractors, office memos and humans. It also includes systems which traditionally we think of as social or human-dominated, such as religious institutions or NGOs." (Haff, 2014a). While Haff capitalises on a geophysical perspective on the dynamics of large-scale technological systems, he still includes social-dominated systems into his picture of the technosphere. To develop our arguments further, we attempt to distinguish more clearly between those two classes of phenomena that are emergent from the point of view of human individuals and technological objects: (i) macrosocial entities and structures such as social networks, governments and bureaucracies, religious institutions or non-governmental institutions (NGOs) and (ii) technological macrostructures such as the internet or large-scale energy and resource extraction and transport systems. While such a classification is not always strictly feasible due to the myriad of interdependencies and co-enabling effects in densely entangled social and technological macrosystems, it is useful for distinguishing agency on the level of human individuals with respect to macrosocial entities and structures from the *macro-agency* of social macrostructures with respect to technological macrostructures. We refer to macro-agency as the *collective agency* of social macrostructures in the sense of their capability to govern, influence, direct and transform technological macrostructures. It should be stressed that this macro-agency arises from the individual agencies and is not an expression of an independent will, it is an emergent macro-phenomenon of networked individuals. Macro-agency differs qualitatively from the agency of human individuals because it is subject to distinct and strong path dependencies and self-set rules.

# Representation of technological systems in Earth system modelling

In Earth system analysis, mathematical and computer models are used as the main analytical tool to gain insights into the functioning and future development of components of the Earth system and of the system as a whole. However, the representation of human societies and technology pose great challenges to formal modelling. Human activities as a whole are modelled in a number of different ways at several scales (Verburg et al., 2016). At present, most global models such as those employed in the assessment reports of the Intergovernmental Panel on Climate Change (IPCC) do not do an adequate job of simulating the human component of the Earth system in a dynamical way. Most global-level representations are based on general equilibrium models of the economy, which often do not include non-linear dynamics (e.g., feedbacks and emergent properties from agent interactions) and are based on strong assumptions about aggregate economic behaviour. For example, integrated assessment models typically only couple biophysical Earth system models (normally climate models) with economic models in a simple, one-way direction (van Vuuren et al., 2012). On the other hand, complex system approaches, such as agent-based models and simple conceptual (toy) models, generally do not operate at the large regional or global levels.

Perhaps an exception to this assessment, while still lacking representations of emergent social-technological structures and dynamics, is the World3 model, made famous by its use in the Limits to Growth scenarios (Meadows et al., 1972). The World3 model is basically a systems dynamics model that is organised around five sectors – population, capital, agriculture, non-renewable resources and persistent pollution (Costanza et al., 2007a). So although it does not contain an explicit technosphere module, World3 does simulate the metabolism of the technosphere – that is, the human commandeering of energy and resources and the expulsion of pollutants into the Earth system – and some of the critical feedbacks associated with this metabolism. Importantly, the model describes the metabolism of the technosphere as a deterministic dynamical system without invoking explicit representations of the agency of a social planner seeking optimal trajectories according to some prescribed utility function. Intriguingly, World3 does a remarkably good job of simulating the observed metabolism of the technosphere from the early 1970s to the present (Turner, 2014).

An early attempt at building a simple conceptual model of the technosphere itself, particularly its internal structure and dynamics, arose from an analysis of the dynamics of the post-1950 Great Acceleration (Figure 1; Hibbard et al., 2006; Steffen et al., 2007; 2015a). Although developed before the concept of the technosphere was published, this simple conceptual model has several features that are consistent with the technosphere idea and thus may provide a starting point for including it in simple World-Earth system models that represent the coevolutionary dynamics of social-technological macrostructures ('World') and biogeophysical processes ('Earth'). First, the core of the model is a production/consumption loop, driven by energy, which can be linked to a biophysical Earth system model via resource use and waste output. Second, the critical role of science, technology and knowledge (which can include cultural norms and values) in driving the production/consumption loop is explicitly included. Third, the role of human agency via institutions and political economy is included at a scale consistent with the technosphere concept. Although a very simple conceptualisation, this model describes ".. a human-created system…that operates beyond our control and that imposes its own requirements on human behaviour" (Haff, 2014a).

Haff's technosphere concept raises important questions about the adequate representation of social and technological mechanisms and constraints in Earth system models. It presents (at least) three basic challenges for current approaches to Earth system modelling:

- the technosphere's internal complex dynamics – feedbacks, networks, emergent properties (Verburg et al., 2016) – must be simulated at the global level;

- it must be interactively coupled with the rest of the Earth system at the appropriate scale, and its most basic metabolic interactions – the commandeering of energy and resources and the expulsion of waste materials (pollutants) back into the rest of the Earth system -  must be simulated; and

- the model must account for human (macro-) agency at the appropriate organisational and spatial scale, implying for instance that individual humans cannot influence the technosphere at the scale that matters for Earth system dynamics.

In the following, we discuss adaptive coevolutionary modelling approaches, which might help to tackle these challenges.

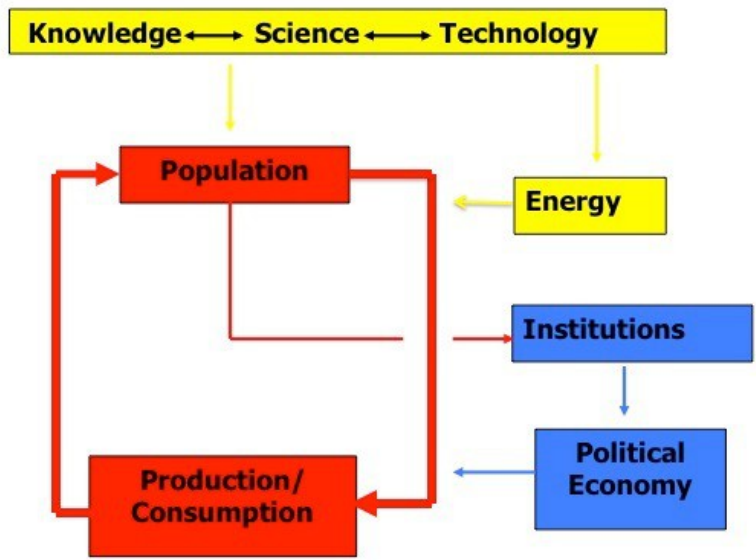

*Figure 1*. An early conceptual model of the technosphere based on an analysis of the dynamics of the Great Acceleration. Adapted from Hibbard et al. (2006).

## The technosphere and emergence of complex coevolutionary dynamics

The technosphere can be thought of as an emergent, coevolved phenomenon of human societies. Issues of scale interaction between the technosphere and human societies should therefore be considered from a coevolutionary perspective. Understanding the emergence of the processes and pathways involved can shed light into the nature of today's interactions between the technosphere and its social sphere of origin. This is particularly important when transitioning from diagnostics of historical developments to projections of possible future trajectories.

From the early palaeolithic, human societies have been characterised by an interwoven complex of technological practices and non-trivial social structures (Camps and Chauhan, 2009). Technologies shaped social structures, while social structures governed the use of technologies (Boserup, 1965). The post-glacial transition from the mesolithic to the neolithic is best understood as a transition in a socio-technological complex (Weisdorf,

2005). Early civilizations were enabled by the technologies they produced and in turn structured by the demands of maintaining these technologies in villages, cities and subsequently across empires (Tainter, 1990). Today, a techno-industrial complex is producing wide-ranging social consequences from the structure of the cities we live in to the channels of communication we use to the daily journeys we undertake. In turn, social dynamics and the resulting systems of preference are continuously influencing the directions and forms technological systems take.

In both cases, the technological and the social, history has seen an emergence of interrelated macro-scale structures. The Great Acceleration of the post-war era should be seen not only as a marked acceleration of the environmental impacts of industrialisation, technological innovation, increased global connectivity, availability of energy and the break-through of globalised neoliberal market principles against imperial divisions of territories and practices (Costanza et al., 2007b). It should be equally seen as the more substantial emergence of increasingly large-scale and complex global technological and social structures, namely the technosphere and the human sociosphere (where the word "sphere" denotes planetary-scale effects). The key question regarding the position of the technosphere in this coevolutionary emergence, today leading to an impending environmental overexploitation of the Earth system with potentially undesirable or even catastrophic outcomes for human societies, is that of collective agency of human societies over the technosphere, as outlined by Haff (2014a).

From the viewpoint of coevolutionary emergence, at issue is both the relationship between agency of individual humans vis-a-vis their social macrostructures such as international institutions, industrial complexes and bureaucratic states, and the collective macro-agency of these macro-scale social entities vis-a-vis the technological macro-infrastructures they collectively have produced and set on their trajectories (Figure 2). Again from a coevolutionary viewpoint, social macrostructures are the product of evolved networks of social interactions. Equally, the technosphere can be conceived of as a network of evolved technological interdependencies, resource and information flows, actions by individuals induced by the technological systems, and interactions with social macrostructures. Haff's particular question on the technosphere concerns the physical and chemical laws governing technological macro-systems. However, since macro-social and macro-technological complexes have coevolved, there is also a large number of interconnections between the social and technological realms that govern their joint trajectories. The dependencies do not run largely in one direction, from the technological to the social, as Haff implies. Rather, the open question encountered is that concerning joint coevolution, that is, the nature of the coupled interplay between social and technological dynamics.

From the viewpoint of the individual, the challenge is twofold: understanding the relationship of individual agency vis-a-vis macrosocial structures, i.e. the role of the individual as part of an increasingly interconnected mega-society and its institutions, and the interrelationship of these mega-societies with their technocomplexes (Figure 2). In all of this, one should keep in mind that technological realities are heterogeneous across the globe, that historical evolution is spatially asynchronous and shaped by regional preconditions, cultures and preferences. Nonetheless, the present-day dominance of industrialisation following a Western development model is striking and seems to be, at least in the present, an attractor of the socio-technological complex once it has emerged.

## Modelling the technosphere as adaptive social-technological-ecological networks

To study the technosphere therefore requires three ingredients: consideration of coevolution and emergence, consideration of social, technological and environmental-ecological networks and their coupled macro-dynamics, and considerations of complexity in these dynamics. Only when this tangle of coevolutionary effects is somewhat understood would the tools be at hand to ask once more questions about the extent and particular role of human agency in governing the technosphere. To assume that a decoupling of scales occurred between the social and technological macro-levels and the level of individual agency is to downplay the collective effects of a multitude of networks that link the scales. These networks, transferring agency, if indirectly, produce feedbacks between the scales, the overall dynamics of which are hard to predict without the aid of systematic, methodologically sound modelling of complex networks.

We suggest that computer simulation models of the technosphere in an Earth system context as an intertwined social-technological-ecological system should be formulated as adaptive network models (Figure 2; Gross and Blasius, 2008; Gross and Sayama, 2009). These models contain at their core an explicit representation of the coevolutionary dynamics of the states of social, technological and ecological entities (nodes) and their connectivities and interdependencies (links). Within the framework of adaptive coevolutionary networks, social processes such as opinion, preference and coalition formation (Holme and Newman, 2006; Auer et al., 2015; Wiedermann et al., 2015) can be integrated with the metabolic network dynamics of technological infrastructures (Bettencourt et al., 2007; Jarvis et al., 2015) and technological change and innovation, none of which are represented in state-of-the-art Earth system or mainstream integrated assessment models. This perspective is in line with, and should integrate, efforts to apply complex systems approaches and agent-based modelling techniques to the study of the economy (Farmer and Foley, 2009; Farmer et al., 2015) as a key constituent of the technosphere. In such an adaptive network modelling system, human agency would be reflected through decision rules and strategies implemented at different levels of social hierarchy and coarse-graining. The effectiveness of this agency would then be revealed by the degree of their manifestation in the structures and dynamics emerging on macroscopic scales (Figure 2).

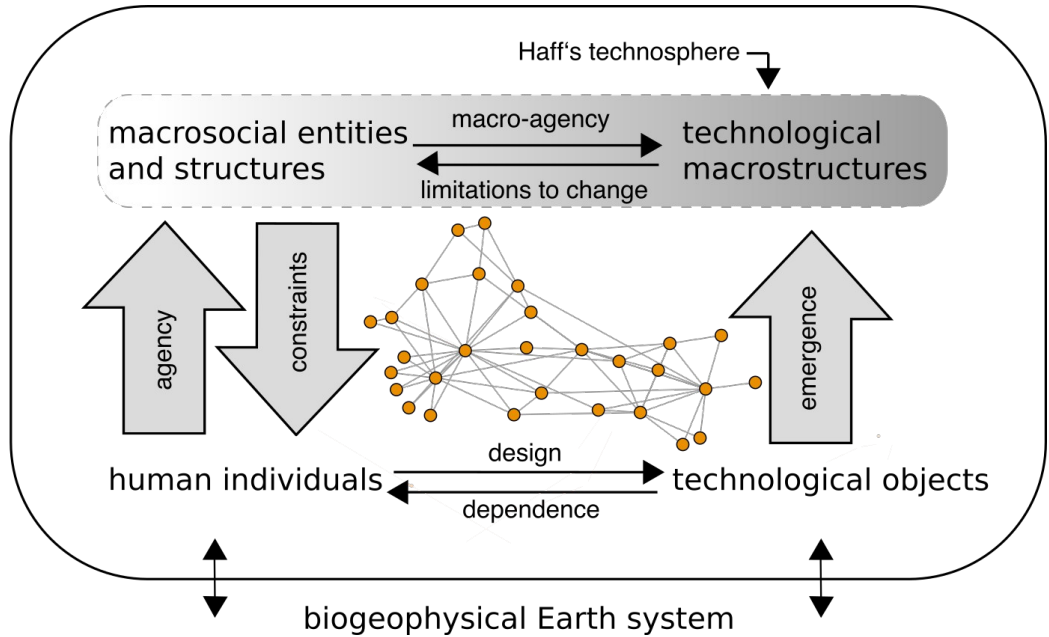

*Figure 2*. The technosphere reconceptualised as an emergent phenomenon in adaptive social-technological networks in the World-Earth system. The figure illustrates the distinction between individual human agency (micro-level) with respect to influencing macrosocial entities and structures (e.g., nation states, bureaucracies and other social institutions.) and their collective macro-agency with respect to technological macrostructures (e.g., the internet, global energy system, industrialised food production). In contrast, Haff's technosphere concept capitalizes on individual human agency mainly with respect to technological macrostructures, but also social macrostructures.

Such a modelling effort would need an enormous amount of data and the theoretical knowledge to make use of it, regarding for example the drivers of technological development, government and business decision making on resource use and emissions, and preference formation of consumers (Helbing et al., 2012; Verburg et al., 2016). A big challenge will be to integrate social science research, that operates in case-specific contexts, with the generalizing framework of Earth system models. In view of computational limitations, such models will only work by making significantly simplifying assumptions and generalizations about the complex dynamics of the Earth system, the social metabolisms that operate within it and the environmental and social feedbacks between them. Therefore, we want to stress that the modelling of social-technological systems and, hence, the technosphere, should not aim primarily at prediction of single future development pathways, but at increasing the understanding of their macroscopic properties and emergent dynamics. Such properties of interest include (i) the coarse-grained topology of World-Earth system state space regions of qualitatively different degrees of desirability and safety (including safe and just operating spaces) and resulting management dilemmas (Heitzig et al., 2016); (ii) critical control points for the technosphere where human agency can trigger transformative change, e.g. in the energy system; and (iii) interactions between social-technological and climatic tipping processes (Schellnhuber, 2009). In this context, it will be relevant to deal with the fact that the self-referentiality of the modeller herself and the infrastructures supporting modelling are parts of the system (the technosphere /Earth system) that she is trying to model. This analytical complication is related to the progression from first order to second order geocybernetics in Earth system analysis as discussed by Schellnhuber (1998).

## Conclusions

Reflecting on Haff's technosphere from the point of view of Earth system analysis, we argue on the one hand that in discourses on sustainable development and global change it is highly relevant to take into account explicitly the constraints imposed on human actions by the technosphere (e.g. intrinsic inertia of technological systems), as well as unanticipated risks resulting from feedback dynamics. In addition to environmental risks related to the transgression of planetary boundaries, examples for unpredictable human extinction-level hazards (and related environmental impacts) associated with technological advances including biotechnologies and the emergence of general artificial intelligence (related to the concept of the *singularity*; Bostrom, 2014) are increasingly coming into the focus of scientific scrutiny as reflected, e.g. by the recent formation of the University of Cambridge Centre for Study for Existential Risks (http://cser.org/). On the other hand, emergent dynamics of the technosphere do not necessarily imply extensive loss of human (macro-) agency as arguably exemplified by the German Energiewende, planned decarbonisation policies in the wake of the Paris 2015 climate agreement, and the social movement on divestment from fossil fuels (Schellnhuber et al., 2016). Consequently, the technosphere should be studied as a coevolutionary planetary phenomenon that can be understood by means of complex systems theory. Computer simulation models as the prominent tools of Earth system analysis play a major role in this endeavour. Therefore, the dynamics of the technosphere and networked feedback processes with the human socio-cultural sphere and the biogeophysical environment need to be captured in next generation models, World-Earth models, to paint a comprehensive panorama of global sustainability. By allowing a focus on highly relevant emergent critical phenomena such as social-technological tipping elements and their interactions with climatic and biospheric tipping processes, such analytical tools can provide a novel and much needed systemic perspective on the safe and just operating space for humanity and can characterise transformative pathways that lead towards it.

## Acknowledgements

This paper was developed in the context of project COPAN on Coevolutionary Pathways in the Earth system at the Potsdam Institute for Climate Impact Research (www.pik-potsdam.de/copan). We acknowledge the attendants of the Technosphere-Coevolution working group at the Anthropocene Curriculum in November 2014 organized by the Haus der Kulturen der Welt and the Max Planck Institute for the History of Science, Berlin, Germany as well as the participants of the LOOPS 2015 workshop on "From Limits to Growth to Planetary Boundaries: Defining the safe and just space for humanity" in Southampton, UK for valuable discussions.

## Funding statement

J.F.D. is grateful to the Stordalen Foundation (via the Planetary Boundary Research Network PB.net) and the Earth League's EarthDoc program for providing financial support. F.M.-H. acknowledges funding by the DFG (IRTG 1740/TRP 2011/50151-0).